\documentclass[a4paper,final]{article}
\usepackage{latexsym,amsmath,amssymb,showkeys,psfig}

\newcommand{\scri}{\mathcal{J}}

\newcommand{\cC}{\mathcal{C}}

\newcommand{\cL}{\mathcal{L}}

\newcommand{\cT}{\mathcal{T}}

\newcommand{\eps}{\epsilon}
\newcommand{\del}{\nabla}

\newcommand{\scalprod}[2]{{\langle #1|#2\rangle}}

\newtheorem{theorem}{Theorem}[section]

\numberwithin{equation}{section}

\title{Calculating initial data for the conformal Einstein equations
  by pseudo-spectral methods}
\author{J\"org Frauendiener\\
Institut f\"ur Theoretische Astrophysik,\\
Universit\"at T\"ubingen,\\
Auf der Morgenstelle 10,\\
D-72076 T\"ubingen,\\
Germany}

\begin{document}
\maketitle

\begin{abstract}
We present a numerical scheme for determining hyperboloidal initial
data sets for the conformal field equations by using pseudo-spectral
methods. This problem is split into two parts. The first step is the
determination of a suitable conformal factor which transforms from an
initial data set in physical space-time to a hyperboloidal
hypersurface in the ambient conformal manifold. This is achieved by
solving the Yamabe equation, a non-linear second order equation. The
second step is a division by the conformal factor of certain fields
which vanish on $\scri$, the zero set of the conformal factor. The
challenge there is to numerically obtain a smooth quotient. Both parts 
are treated by pseudo-spectral methods. The non-linear equation is
solved iteratively while the division problem is treated by
transforming the problem to the coefficient space, solving it there by
the QR-factorisation of a suitable matrix, and then transforming back.
These hyperboloidal initial data can be used to generate general
relativistic space-times by evolution with the conformal field
equations. 
\end{abstract}

\section{Introduction}
\label{sec:intro}

In this article we shall discuss the problem of numerically
calculating ``hyperboloidal initial data''. These occur naturally in
the numerical solution of the conformal Einstein equations, a promising
approach towards the numerical evolution of general relativistic
space-times \cite{jf-1997-2,jf-1997-3,Huebner-1998-2,Huebner-1995}.

Consider an asymptotically flat solution of the Einstein equation. Let
$\Sigma$ be a space-like hypersurface which extends to infinity in
such a way that it reaches null infinity $\scri$. Assume that it
remains space-like even in the limit then it will touch $\scri$
transversely. Prototypes for such hypersurfaces are the space-like
hyperboloids in Minkowski space, hence they are called hyperboloidal
hypersurfaces. These are rather different from the standard
asymptotically euclidean hypersurfaces which end up at space-like
infinity $i^0$.  In contrast to the latter hypersurfaces the
hyperboloidal hypersurfaces are not Cauchy hypersurfaces of the
complete space-time for the standard Einstein evolution. However, they
can still be used for pre- or retrodiction depending on whether they
end up on $\scri^+$ or $\scri^-$.

The main reason for focusing on hyperboloidal hypersurfaces is the
fact that outgoing radiation can be traced much more effectively on
those hypersurfaces than it is possible on the asymptotically flat
hypersurfaces \cite{Huebner-1998-2}. This is due to the fact that they
approximate a null hypersurface at large distances along which the
radiation propagates instantaneously.

The initial data for the Einstein evolution on a hyperboloidal
hypersurface $\Sigma$ implied by the space-time geometry consists of
the intrinsic metric and the extrinsic curvature. The requirement that
$\Sigma$ should reach out to null infinity imposes quite definite
asymptotic fall-off conditions on these fields on $\Sigma$. After a
suitable conformal compactification of the space-time the fall-off
conditions can conveniently be captured in smoothness conditions for
appropriately rescaled fields on the boundary of a three-dimensional
manifold. In this ``conformal'' picture the evolution of initial data
is most appropriately performed with the regular conformal field
equations \cite{Friedrich-1981,Friedrich-1983}, yielding the
``conformal method'' for solving Einstein's equations. The main
advantage of using this method is that the conformal field equations
allow the evolution of initial data all the way up to time-like
infinity $i^+$, which is a regular point of the conformal manifold
provided the data are sufficiently close to Minkowski data. But even
if $i^+$ does not exist one can follow the evolution up to the point
where singularities form. This has been demonstrated
in~\cite{Huebner-1995}. Thus, it is possible to evolve the complete
future of an initial hypersurface on a single grid including the
points of $\scri$ (and even beyond, since one can smoothly extend the
initial data across $\scri$). This allows physically meaningful
quantities like the Bondi-news and -momentum, the radiation flux
etc. which well defined only on $\scri$ to be determined there
without any further approximation, essentially be reading off the
appropriate functions from the grid.

Of course, for the conformal field equations, initial data have to be
determined, too. They consist no longer only of the first and second
fundamental form, satisfying the standard constraint equations. The
conformal field equations comprise more variables than the standard
Einstein equations. In particular, they include the Bianchi identity
from which follow evolution equations for both the Weyl- and the Ricci
curvature. Hence, an initial data set for the evolution with the
conformal field equations is larger than for the standard
evolution. Such an initial data set is called hyperboloidal if the
initial surface is a hyperboloidal hypersurface.

The conformal constraint equations, i.e., that part of the conformal
field equations which is intrinsic to the initial hypersurface, need
to be solved in order to obtain the initial data. At the moment there
is no way to solve those constraints directly in more than one space
dimension but there is a procedure due to Andersson, Chru\'sciel and
Friedrich \cite{AnderssonChruscielFriedrich-1992} which allows the
construction of hyperboloidal initial data. Essentially, one solves
one non-linear second order equation for one scalar function to
determine a conformal factor which transforms from a solution of the
constraints in physical space to unphysical space. Once this has been
found one can determine the remaining hyperboloidal initial data
algebraically from a part of the conformal constraint equations.

In this procedure there are two complications. In the first place, the
second order equation is singular at the boundary where the
hyperboloidal surface intersects $\scri$. Fortunately, this turns out
to be more trouble for the analytical than for the numerical
treatment. And in the second place, when computing the initial data
one has to divide by the conformal factor which vanishes on
$\scri$. Although it can be shown analytically that this ``division by 
zero'' is well defined it does pose numerical problems.

The purpose of this article is to discuss a numerical implementation
for finding hyperboloidal initial data by pseudo-spectral methods and
to suggest one possible way to overcome the division problem. In
section \ref{sec:hyp}. we describe the analytical background in more
detail. The numerical solution of the second order equation is
discussed in section \ref{sec:yam} while the division problem is
treated in section~\ref{sec:div}. A short discussion concludes the
paper.

\section{Finding hyperboloidal initial data sets}
\label{sec:hyp}

Let $(\tilde M,\tilde g)$ be an asymptotically flat solution of the
vacuum Einstein equation which has a smooth conformal structure at
null infinity $\scri$. Consider $\tilde\Sigma$, a hyperboloidal
hypersurface intersecting $\scri$ in a smooth two-dimensional surface
$\del \Sigma$. The Lorentz metric $\tilde g$ on $\tilde M$ induces a
riemannian metric $\tilde h$ on $\tilde \Sigma$. Let $\Omega$ be a
conformal factor, which smoothly attaches null infinity to $\tilde M$
thus defining the ``conformal'' space $M=\tilde M \cup \del M$. Then
it follows that $\Sigma = \tilde\Sigma \cup \del\Sigma$ is a smooth
submanifold with boundary in $M$ and that $\Omega > 0$ on
$\tilde\Sigma$ and $\Omega=0$, $d\Omega\ne0$ on $\del
\Sigma$. Furthermore, there exists a smooth riemannian metric $h$ on
$\Sigma$ such that on $\tilde \Sigma$ the relation
\begin{equation}
  \label{2.1}
  h = \Omega^2\tilde h
\end{equation}
holds. The embedding of $\tilde\Sigma$ in $\tilde M$ defines the
second fundamental form $\tilde k$ on $\tilde \Sigma$. Being induced
from a vacuum solution of the Einstein equation the pair $(\tilde h,
\tilde k)$ satisfies the physical constraint equations on $\tilde
\Sigma$.

When solving the constraint equations on asymptotically euclidean
initial surfaces it is convenient to make the assumption of
time-symmetry, namely that the extrinsic curvature of the initial
surface should vanish. This condition is not compatible with the
geometry of a hyperboloidal surface but one can obtain similar
simplications in this case also by the assumption that the extrinsic
curvature be pure trace, i.e., proportional to the metric,
\begin{equation}
  \label{2.2}
  \tilde k = \frac13 \tilde K \tilde h.
\end{equation}
Then the momentum constraint implies that $\tilde K$ is a constant and
therefore, so is the scalar curvature ${}^3\tilde R$ of $\tilde h$ by
virtue of the hamiltonian constraint.  One can assume that (after
rescaling $\tilde h$ with a suitable constant)
\begin{equation}
  \label{2.3}
   {}^3\tilde R = -6.
\end{equation}
Note, that the assumption \eqref{2.2} which asserts that the tracefree
part of the extrinsic curvature vanishes is conformally invariant so
that the extrinsic curvature of $\Sigma$ in the unphysical space is
also pure trace, albeit not necessarily constant.

In this paper we will impose the condition \eqref{2.2}. This is yields
simplest case for constructing hyperboloidal initial data sets. The
analytic treatment of this problem has ben thoroughly discussed in
\cite{AnderssonChruscielFriedrich-1992}. But there are also several
other, less restrictive, treatments in the literature. In
\cite{AnderssonChrusciel-1994} the assumption \eqref{2.2} is dropped
allowing for an extrinsic curvature which is almost general apart from
the fact that the mean curvature is required to be constant. In
\cite{IsenbergPark-1996} also this requirement is dropped and in
\cite{Kannar-1996} the existence of hyperboloidal initial data is
discussed for situations with a non-vanishing cosmological constant.

In order to construct hyperboloidal initial data one may proceed as
follows \cite{Friedrich-1993,AnderssonChruscielFriedrich-1992}: let
$\omega$ be a boundary defining function for $\Sigma$, i.e., $\omega >
0$ on $\Sigma$ and $\omega=0$, $d\omega\ne 0$ on $\del\Sigma$ and let
$h$ be a smooth metric on $\Sigma$. Now we seek a conformal factor
$\Omega$ so that the metric $\tilde h=\Omega^{-2} h$ defined on
$\tilde\Sigma$ satisfies \eqref{2.3}. We write $\Omega =
\phi^{-2}\omega$ for some smooth function $\phi$ on $\Sigma$. Then
\eqref{2.3} turns into the non-linear second order equation, also
called the Yamabe equation
\begin{equation}
  \label{2.4}
  4 \omega^2 \Delta \phi - 4 \omega \del^a\omega \del_a
    \phi - \left[ \omega^2 R + 2 \omega \Delta \omega -3
    \del^a\omega\del_a \omega \right] \phi = 3 \phi^5. 
\end{equation}
Here, we have used the Laplace operator $\Delta = \del^a\del_a$ with
respect to $h$, the covariant derivative operator $\del$ and the
scalar curvature $R$ of $h$. The most obvious property of this
equation is the fact that it degenerates on the boundary where
$\omega$ vanishes. This is not entirely surprising considering its
origin: if the equation were regular on the boundary one would
presumably be able to specify boundary data, thus introducing some
freedom into the structure of a hyperboloidal hypersurface at its
intersection with $\scri$. But this would be in contradiction with the
fact that $\scri$ is universal, being fixed entirely by the smooth
conformal structure. The conformal transfomation properties of
\eqref{2.4} ensure that the conformal factor $\Omega$ defined from a
solution $\phi$ does not depend on the specific form of the boundary
defining function $\omega$ and, furthermore, that it depends only on
the conformal class of the metric $h$.

Consider now the following fields defined on $\Sigma$
\begin{align}
\label{2.5}
\Phi_{ab} & = - \Omega^{-1} \left(\del_a\del_b \Omega - \frac13
  h_{ab} \Delta \Omega\right),\\
  \label{2.6}
  E_{ab} &= \Omega^{-1} \left(R_{ab} - \frac13 h_{ab} R -
    \Phi_{ab}\right).
\end{align}
These fields are the essential initial data necessary for the
evolution with the conformal field equations. $\Phi_{ab}$ is the
projection of the conformal Ricci tensor onto $\Sigma$, while
$E_{ab}$ represents the rescaled electric part of the Weyl tensor.  As
they stand these expressions are valid only on $\tilde\Sigma$, being
formally singular on the boundary where $\Omega$ vanishes and one
needs to worry whether there exists a smooth extension to
$\del\Sigma$. This question and the more immediate question of
existence, uniqueness and regularity of solutions of \eqref{2.4} have
been answered in complete detail in
\cite{AnderssonChruscielFriedrich-1992} where the following theorem
has been proved.
\begin{theorem}
\label{theo:1}
Suppose $(\Sigma,h)$ is a three-dimensional, orientable, compact, smooth
Riemannian manifold with boundary $\del\Sigma$. Then there exists a
unique solution $\phi$ of \eqref{2.4} and the following conditions are
equivalent:
\begin{enumerate}
\item The function $\phi$ as well as the tensor fields \eqref{2.5} and
  \eqref{2.6} determined on $\tilde\Sigma$ from $h$ and
  $\Omega=\phi^{-2}\omega$ extend smoothly to all of $\Sigma$.
\item The conformal Weyl tensor $C_{ab} = \Omega E_{ab}$ goes to
  zero on $\del \Sigma$.
\item The conformal class of $h$ is such that the extrinsic curvature
  of $\del\Sigma$ with respect to its embedding in $(\Sigma,h)$ is
  pure trace.
\end{enumerate}
\end{theorem}
Condition 3. is a weak restriction of the conformal class of the
metric $h$ on $\Sigma$, since it is only effective on the boundary.
Interestingly, the theorem only requires $\Sigma$ to be orientable and
does not restrict the topology of $\Sigma$ any further.  We exploit
this fact by assuming the existence of an isometric and hypersurface
orthogonal action of $U(1)$ without fixed points on $\Sigma$ which we
take to have topology $S^1\times S^1 \times I$. Thus, we may ignore
the dependence on one coordinate, reducing the problem to one on the
two-dimensional surface $S^1\times I$. We take coordinates on this
surface as $(u,v) \in [0,2\pi] \times [-1,1]$, with $2\pi$-periodicity
implied on the $u$-dependence. The boundary is given by $v=\pm 1$. In
the case treated here in detail, we choose $\scri$ to coincide with
the boundary. We refer to \cite{Schmidt-1996} for a more detailed
discussion of space-times with these properties. In more complex cases
one could think of choosing boundary functions $\omega$ which vanish
not only on the boundary but also somewhere in the interior. This
would define and evolve to physically interesting space-times with
more complex geometries \cite{Huebner-1998-2}. We will show one
possibility below.

We choose the boundary function to be  
\[ 
\omega(u,v)=\frac12\left(1-v^2\right).
\]
On the boundary the equation \eqref{2.4} reduces to
\begin{equation}
  \label{2.6-1}
\left(\del^a\omega \del_a\omega\right)\,\phi=\phi^5.  
\end{equation}
Since $\phi$ has to be non-zero on the boundary this implies $\phi =
\sqrt[4]{\del^a\omega \del_a\omega}$, so that the boundary values of
the solution are completely fixed by the equation.

Due to our assumptions the metric on $\Sigma$ has the form
\[
h=h_{uu}\,du^2 + 2h_{uv}\,du\,dv+ h_{vv}\,dv^2 + h_{ww}\,dw^2,
\]
where the metric functions do not depend on the coordinate $w$. Since
we need to specify only the conformal class of a metric on $\Sigma$ we
may assume that $h_{ww}=1$, leaving the other functions arbitrary
except for the condition 3. of Theorem \ref{theo:1}. With these
assumptions, the induced metric on the boundary is
\[
p = h_{uu}\,du^2 +dw^2,
\]
and the extrinsic curvature of the boundary with respect to the metric 
$h$ is proportional to
\[
\lambda = \left(h^{uv}h_{uu,u} + h^{vv}h_{uu,v} + 2 h_{uu}
  h^{uv}{}_{,u} \right)\,du^2.
\]
Condition 3. requires that $\lambda$ be proportional to the induced
metric $p$ which implies, that $\lambda$ itself has to vanish. One
possibility to satisfy this condition is to require that $h_{uv}=0$
and $h_{uu,v}=0$ on the boundary. It is worthwhile to point out again
that the purpose of condition 3 is to ensure the smooth extensibility
of the solution and the tensor fields mentioned in
theorem~\ref{theo:1}. It is possible to find solutions of \eqref{2.4}
for free data which do not satisfy condition~3.

\section{Numerical solution of the Yamabe equation}
\label{sec:yam}

We have implemented a numerical scheme for the solution of \eqref{2.4}
based on pseudo-spectral methods. 

In this section we want to describe a numerical scheme based on
pseudo-spectral methods for solving the Yamabe equation.  There are
various reasons for considering these methods. They are known for
their high accuracy, at least in situations where the solution is
smooth. Then the numerical error decreases exponentially with the
number of degrees of freedom (i.e., the number of collocation points
or the number of basis functions used to approximate the
solution). This is much faster than the error decay in any finite
difference method which is $O(N^{-q})$ with $q$ usually less than
$4$. This property makes pseudo-spectral methods ideally suited for
elliptic problems.  Pseudospectral methods have been employed
successfully in various areas of physics and applied mathematics. In
particular, we mention the work of S. Bonazzola and his co-workers
e.g., \cite{BonazzolaEtAl-1998} on various applications in
relativistic astrophysics.

Let us briefly describe the basic idea behind the use of spectral and
pseudo-spectral methods.  For more information on these methods we
refer to the standard textbooks
\cite{CanutoEtAl-1988,Fornberg-1996,GottliebOrszag-1977}. When solving
a partial differential equation (time independent for our immediate
purposes) in some domain $\Sigma$
\begin{equation}
  \cL f = 0
\end{equation}
where $\cL$ is some non-linear differential operator one seeks a
solution in the form of an expansion in some suitable basis functions
(assumed to be a complete set on the region of interest)
\begin{equation}
  f(x)= \sum_{n=1}^N f_k \phi_k(x).
\end{equation}
The most popular functions in use are the trigonometric polynomials $e^{imx}$
and the Chebyshev polynomials $T_m(x)$. The choice depends on the
topology of the domain and the boundary conditions. Inserting this
expansion into the PDE yields a system of equations for the expansion
coefficients. There are various ways to set up these
equations. Spectral methods such as the Galerkin method reexpand
$\cL(\sum_{n=1}^N f_k \phi_k(x))$ in terms of the basis
functions. This is only realistic in a few cases, mostly if $\cL$ is
constant and linear. In most other cases, the determination of the
expansion coefficients of $\cL f$ in terms of those of $f$ is either
impossible or computationally too expensive. Then one can fall back on
the pseudo-spectral or collocation method: one introduces suitable
collocation points $x_1,\ldots,x_N$ and then the approximate solution
is forced to satisfy the equation at the inner points and the boundary
conditions. This yields $N$ equations for the $N$ expansion
coefficients.

The existence of the collocation points allows dual representations of
the function $f$. Besides the ``physical'' representation based on the
function values $f(x_i)$ at the collocation points there is the
``spectral'' representation based on the expansion into basis
functions. The idea of the collocation method is to switch freely
between those two representations using whichever is best to evaluate
the various terms in the operator. This is made possible (at least for
the Fourier and Chebyshev polynomials) by fast Fourier transformation
(FFT) techniques. 

The representation in coefficient space is ideally suited for
efficiently and accurately evaluating derivatives. The coefficients of
the derivative of a function are easily determined from the
coefficients of the function either by simple multiplications or else
by three term recurrence relations. Nonlinear and/or nonconstant terms
in the operator are best computed using the physical representation.

In general, the matrices which represent the spectral approximations
of even the simplest linear operators are full and difficult to invert
directly. Their efficient inversion can be achieved by choosing a
suitable preconditioning operator (see below).

The method we employ for solving the Yamabe equation is an iterative
method using a  defect correction scheme. It is based on the
following observations. We write equation \eqref{2.4} formally as
\begin{equation}
  \label{2.7}
  \cL \phi = \phi^5.
\end{equation}
Here, $\cL$ is a linear operator made up from derivative and
multiplication operators. We construct a Richardson iteration
procedure by writing $\phi^{n+1} = \phi^n + \delta\phi$. Inserting
this into \eqref{2.7} and ignoring terms of higher order than the
first in the correction term $\delta\phi$ yields
\begin{equation}
  \label{2.8}
  \cL\delta\phi - 5\left(\phi^n\right)^4\,\delta\phi = -\left(
  \cL\phi^n - \left(\phi^n\right)^5 \right).
\end{equation}
Thus, the general procedure for solving \eqref{2.4} is the
following. Suppose we have some suitable approximation $\phi^n$ and
compute the residual $r^n=\cL\phi^n - \left(\phi^n\right)^5$. Then
solve the \emph{linear} equation $\left(\cL - 5\left( \phi^n \right)^4
\right)\, \delta\phi=-r^n$ to obtain the correction $\delta\phi$ and
an updated guess $\phi^{n+1} = \phi^n +\delta \phi$. With that repeat
the procedure until the accuracy is satisfactory. We observe that the
linear operator acting on $\delta\phi$ is also updated at each step
but only by diagonal terms.  

As pointed out above, for pseudo-spectral methods, the matrix
representation of the operator $\cL$ is generally a full $N\times N$
matrix $\cL_{PS}$, so that inversion is prohibitive both in terms of
time and storage requirements for high $N$ and especially in higher
dimensions. However, there exists a way to circumvent this problem
which is due to S. Orszag \cite{Orszag-1980}: instead of inverting the
pseudo-spectral representation of $\cL$, we substitute a
finite-difference approximation $\cL_{FD}$ of $\cL$ into the left-hand
side of equation \eqref{2.8} and use this for the iteration
procedure. In general, FD-approximations have sparse matrix
representations, so that the iteration equation can be solved
efficiently. Note, that this substitution is only made on the
left-hand side of the equation while on the right-hand side the full
pseudo-spectral approximation is retained. As pointed out in
\cite{Orszag-1980} this method allows the efficient solution of
general problems with operation costs and storage not much larger than
those of the simplest finite difference approximations to the problem
with the same number of degrees of freedom.

In a sense this is similar to an inexact Newton method where the exact
Jacobian is replaced by some approximation. Under such circumstances
one cannot expect to have the full quadratic convergence of the Newton
method, for which successive errors satisfy $\eps_{n+1} \le K
\eps_n^2$ for some $K>0$. Instead, under most circumstances one can
hope for linear convergence, i.e., $\eps_{n+1} \le K \eps_n$, so that
$\eps_n \le \alpha^n$ for some positive $\alpha<1$ \cite{Kelley-1995}.

The solution procedure outlined in the previous section is implemented 
as follows. The topology suggests that we expand the fields into a
Fourier series in the periodic $u$-direction  and Chebyshev
polynomials in the $v$-direction as in
\begin{equation}
  \label{2.9}
  f(u,v) =  \sum_{m=0}^M \sum_{l=-\frac{N}2}^{\frac{N}{2}-1} f_{lm}
  e^{ilu}\,T_m(v).
\end{equation}
We introduce the collocation points $(u_i,v_k)$, where
$u_i=\frac{2\pi}N i$, $i=0,\ldots,N-1$ and $v_k=\cos(\frac{\pi
k}{M})$, $k=0,\ldots,M$. Then we can switch between the physical and
the spectral representation by using fast transformation
techniques. The free data $h_{uu}$, $h_{uv}$, $h_{vv}$ are specified
on the collocation points subject only to the conditions $h_{uv}=0$
and $h_{uu,v}=0$ on the boundary. From the metric functions the
connection (i.e., the Christoffel symbols) and the scalar curvature
are determined in order to obtain the differential operator $\cL$. As
indicated above, we use both its spectral approximation as well as a
finite difference approximation. In the present case, we take an
approximation which is $2^{nd}$-order accurate. In deriving the
expression for this approximation one has to take into account that
the collocation points $v_k$ are not uniformly distributed.

The equation \eqref{2.8} is then imposed at all the interior
collocation points except at the boundary, where the values for the
solution determined from \eqref{2.6-1} are inserted. This yields a
matrix of size $(N(M-1))^2$, which is sparse. The computation of the
residual, i.e., the right hand side of \eqref{2.8} at each iteration
step is done with the full spectral accuracy. The linear equation for
the correction is solved iteratively by methods taken from the sparse
matrix package {\tt LASPACK} \cite{LASPACK}.

\begin{figure}[htbp]
  \makebox[\hsize]{\hfill\psfig{file=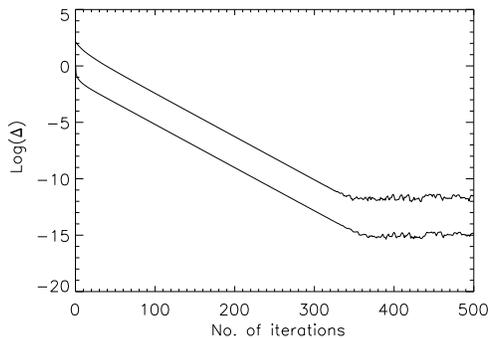,height=5cm}\hfill}
  \caption{Size of residual and of update in the outer loop versus
    the number of iterations. } 
  \label{fig:logerr_n}
\end{figure}

In Fig.~\ref{fig:logerr_n} is shown the convergence behaviour for a
a typical run with 32 degrees of freedom (i.e., base functions) in
each coordinate direction. The upper line is the logarithm (base 10)
of the maximum size of the residual, while the lower line shows the
logarithm of the maximum size of the update at each iteration step of
the outer loop. Consistently, the residual is about three orders of
magnitude above the update. The convergence is exponential according
to the formula
\begin{equation}
  \label{2.10}
  \Delta \propto \alpha^N,
\end{equation}
where we find that in this case $\alpha \approx 0.96$. The value of
$\alpha$ depends on the number of degrees of freedom $M=32\cdot 33$
and the free data. As mentioned above, the convergence behaviour is
not the one expected for a true Newton method. This is not too
surprising since the linear operator we use for obtaining the update
at each iteration step is not the Jacobian of the non-linear
function. When the correction hits the level of machine accuracy no
further reduction in the residual is possible and the convergence
levels off, the residual remaining at a level of about $10^{-12}$. The 
exact numbers depend on the free data.

\begin{figure}[htbp]
  \makebox[\hsize]
  {\hfill\psfig{file=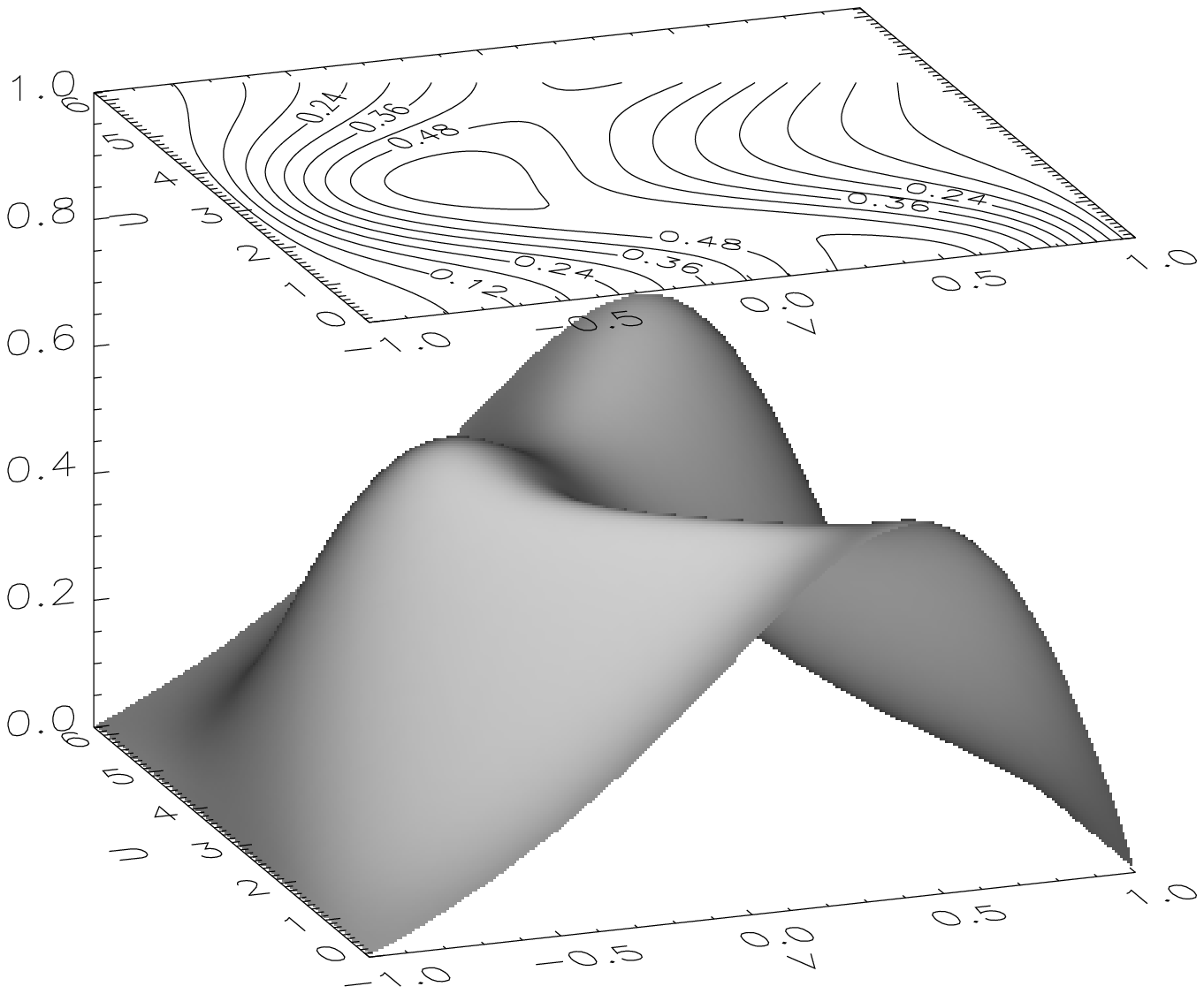,height=5cm}\hfill
    \psfig{file=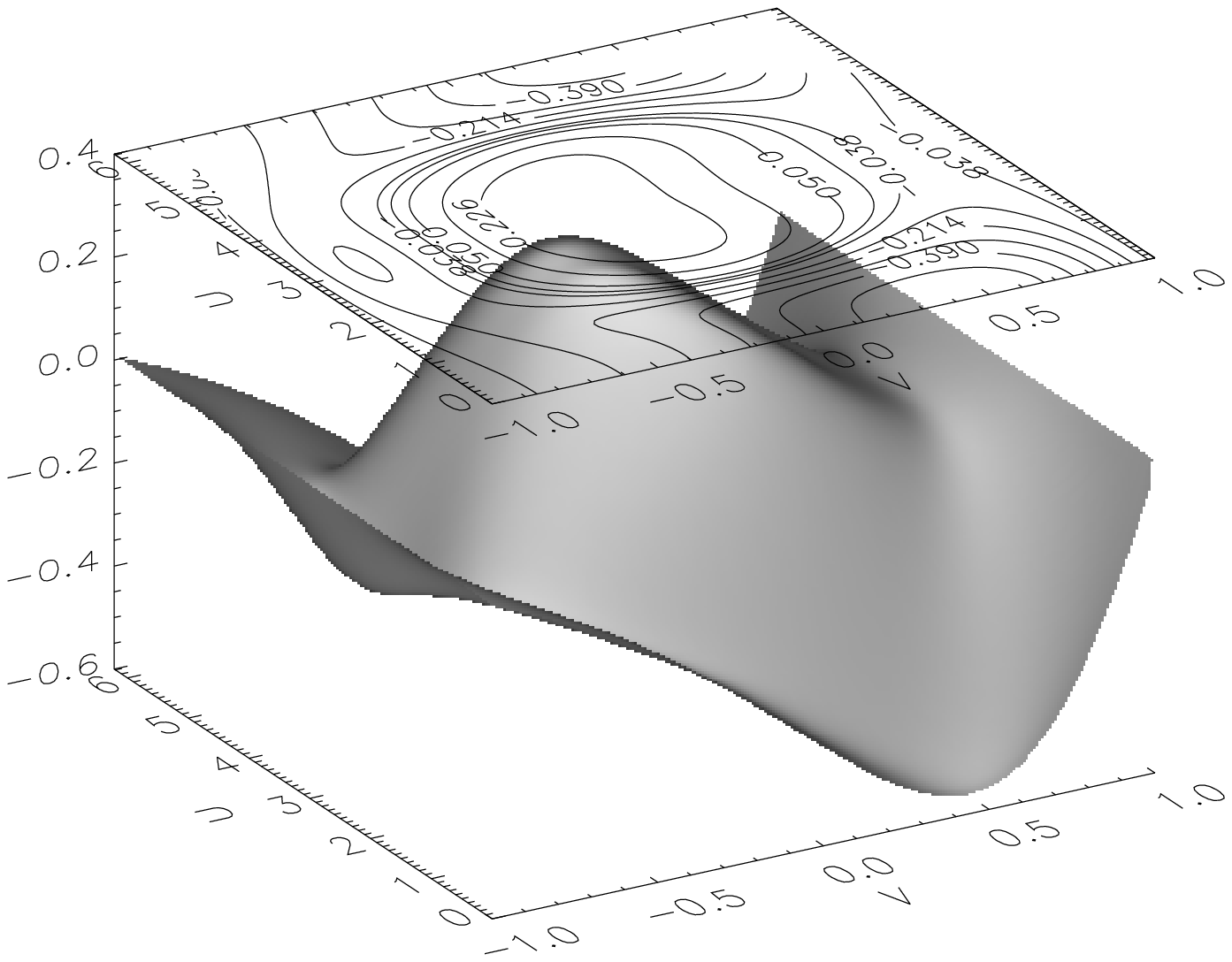,height=5cm}\hfill}
  \parbox{10cm}{\caption{The conformal factor $\Omega$ obtained from the same
    expressions for the metric functions but different boundary
    functions. The region $\Omega\ge0$ corresponds to the physical
    space-time.}} 
  \label{fig:omega}
\end{figure}

In Fig.~\ref{fig:omega} are shown two solutions of the Yamabe equation
for different kinds of free data.
For the left diagram (case 1) we have chosen the following data
\begin{align*}
  \omega &= \frac12\left(1-v^2\right),\\
  h_{uu} &= 2\left( 1+\omega^2\left(v^2-\sin^2u\right)^2\right),\\
  h_{uv} &= 2\omega\left(v^2-\sin^2u\right)\,e^{-v\cos u},\\
  h_{vv} &= 2\,e^{-2v\cos u}.
\end{align*}
The right diagram (case 2) is meant to be a demonstration of the
possibility to specify boundary functions which also vanish inside the 
computational grid. It was obtained by choosing
\[
\omega = \frac12\left(1-v^2\right) \cdot\left(\frac12\left(1+\cos
    u\right)^2 + 2\,v^2-0.7\right)
\]
as the boundary function and keeping the same expressions for the
metric functions. It is apparent from the contour plots that the
different boundary functions correspond to different topologies of the
interior regions. For case~1 there are two $\scri$'s at the grid
boundaries $v=\pm 1$, so that the physical space-time has the topology
$S^1\times I$, as intended. But in case~2, there are additional zeros
of $\omega$ which introduce another $\scri$ inside the grid boundaries
which has a circular topology. The resulting conformal factor defines
the physical region inside that circular $\scri$, giving it the
topology of a disc. This two-dimensional picture is then swept around
by the $U(1)$-action to yield a three-dimensional space-time with the
topology of a full torus whose boundary is $\scri$. This physical
space-time is embedded in an ``unphysical'' region which is again
bounded by a $\scri$ on each side.

We keep the $\scri$'s at the grid boundaries because they prevent us
from having to specify boundary condtions. Choosing the boundary
function appropriately can lead to rather complicated space-times: by
simply changing the sign of the boundary function in the example above
we can switch the interior (physical) and exterior (unphysical)
regions. In this way we obtain a space-time with toroidal $\scri$'s
and another ``asymptotically flat'' end, which one could loosely
interpret as a black hole region. However, one should be careful with
interpretations like this because we still have the assumption of the
existence of a hypersurface orthogonal Killing vector.

\section{The division problem}
\label{sec:div}

We now turn to the problem of calculating the remaining initial data
from a solution of the Yamabe equation. As indicated in
section~\ref{sec:hyp} this involves a division of the tensor
components \eqref{2.5} and \eqref{2.6} by the conformal factor. From
theorem~\ref{theo:1} one knows that these components vanish on $\scri$ 
and that the quotient is smooth across $\scri$. Let us first discuss a
one-dimensional example.

Consider two real valued functions on the interval $[-1,1]$ with $f(0)
= g(0) = 0$ and $f(x)\ne 0$, $g(x) \ne 0$ elsewhere. Our task is to
compute the quotient $f/g$ on the interval. The only problem is at the
origin $x=0$, where the quotient is not defined. Analytically, one can
use l'H\^opital's rule to obtain the limit
$\lim_{x\to0}(f/g)(x)$. Numerically, however, the problem is more
subtle, at least if one is interested in obtaining an answer as
accurately and smoothly as possible.

Let us be more specific in our assumptions on $f$ and $g$. We take
them both to be at least $\cC^2$ and to vanish at $x=0$, but with
$g'(0)\ne 0$. Then we have $f(x)=x\tilde f(x)$ and $g(x) = x\tilde
g(x)$ for $\cC^1$-functions $\tilde f$ and $\tilde g$ with $\tilde
g(0)\ne 0$. Then
\[
\lim_{x\to0}(f/g)(x)=\frac{\tilde f(0)}{\tilde g(0)},
\]
but obviously this limit cannot be calculated numerically in a direct
way. The limit procedure has to be realized somehow. A straightforward 
method would be to approximate 
\[
(f/g)(0)=\frac{f'(0)}{g'(0)} \sim \frac{f(\eps)-f(0)}{g(\eps)-g(0)} =
\frac{f(\eps)}{g(\eps)},
\]
for small enough $\eps$. Unfortunately, this is only a first order
approximation which, of course, could be improved by using more
accurate finite difference formulae for approximating the derivatives
at $x=0$. Still, we do not get the accuracy of a spectral method.

We propose here a method which is more in line with the idea of
spectral methods. Roughly speaking, we transform the problem to
coefficient space, solve it there and then transform back to physical
space. Define $\cT_M=\mathrm{span}(T_m,0\le m\le M)$, the space of
polynomials on $[-1,1]$ with degree at most $M$. This space is spanned
by the first $M+1$ Chebyshev polynomials and it carries a scalar
product defined by
\[
\langle T_m|T_n\rangle = \int^1_{-1}
T_n(x)T_m(x)\,\frac{dx}{\sqrt{1-x^2}} 
\]
The Chebyshev polynomials are orthogonal with respect to this scalar
product, but not normalized. For each $g \in \cT_M$, multiplication by
$g$ defines a linear map $g:\cT_M \to \cT_{2M}$. This map and its
matrix representation with respect to the basis polynomials follow
from the Clebsch-Gordan like formula
\begin{equation}
  \label{4.1}
  T_n\,T_m = \frac12\left(T_{|n-m|} + T_{n+m}\right).
\end{equation}
Thus, e.g., the matrix representation of (multiplication by) $T_0$
is the $(2M+1)\times (M+1)$-matrix which is the identity in the upper half
and zero otherwise, while $T_1$ and $T_2$  have the following $(2M +1)
\times (M+1)$-matrix representations, respectively
\begin{equation}
  \label{4.2}
  \frac12\left(
    \begin{array}{ccccccc}
      &1&&&&&     \\
      2&&1&  &   &&\\
      &1&&1&  &    &\\
      &&1&&1  &    &\\
      &&&\ddots&&\ddots&\\
    \end{array}
  \right),\qquad
  \frac12\left(
    \begin{array}{ccccccc}
      &&1&&&&     \\
      &1&&1&&&     \\
      2&&& &1    &&\\
      &1&&&&1      &\\
      &&\ddots&&&&\ddots\\
    \end{array}
  \right).
\end{equation}
The higher degree polynomials $T_n$ have similar representations. A
general polynomial $g \in \cT_M$ is a linear combination in the
Chebyshev polynomials and hence its matrix representation $\mathbf{G}$
is obtained by adding up these basic representations
appropriately. Having the representation with respect to the basis
polynomials it is easy to obtain the representation with respect to
the normalised Chebyshev  bases in $\cT_M$ and $\cT_{2M}$.

Suppose now that $f$ is in $\cT_{2M}$. Obviously, the image of $\cT_M$
under multiplication by $g$ is an $(M+1)$-dimensional subspace of
$\cT_{2M}$, hence not all $f\in \cT_{2M}$ are also in that image. Our
task is to invert the map $g:\cT_M\to g[\cT_M]$ on its image. Thought
of in terms of linear algebra, this requires to solve $(2M+1)$ linear
equations, of which only $(M+1)$ are linearly independent for $(M+1)$
unknowns. The matrix $\mathbf{G}$ is the coefficient matrix of that
system of equations. 

We can solve this problem by finding the reduced QR-factorisation of
$\mathbf{G}$, i.e., we seek matrices $\mathbf{Q}$ and $\mathbf{R}$ so
that that $\mathbf{G=QR}$, where $\mathbf{R}$ is an upper triangular
$(M+1) \times (M+1)$-matrix and $\mathbf{Q}$ is a $(2M+1)
\times(M+1)$-matrix whose columns are mutually orthonormal. Thus, the
columns of $\mathbf{Q}$ form an orthonormal basis for the image of
$\mathbf{G}$. It should be noted that the scalar product involved here
is the one defined between the Chebyshev polynomials if one uses the
normalised polynomials when representing matrices, which will be
assumed in the sequel. For a more thorough discussion of the
QR-factorisation from various perspectives we refer to standard
textbooks on numerical linear algebra like
\cite{TrefethenBau-1997,GolubVanLoan-1996,StoerBulirsch-1978}.

Suppose now that $\mathbf{G=QR}$ has been factored. We note that
$\mathbf{Q^*Q} = \mathbf{1}_M$, the identity in $\cT_M$, while
$\mathbf{QQ^*}$ is the orthogonal projector onto the image of
$\mathbf{G}$. Given any $f\in\cT_{2M}$, the QR-factorisation enables
us to ``solve'' the overdetermined system $\mathbf{G}h=f$ in the
following way. We have $\mathbf{QQ^*}f = \mathbf{G}h$ for some $h\in
\cT_M$. Hence we get $\mathbf{Q^*}f=\mathbf{R}h$ and, finally,
inverting $\mathbf{R}$, we get $h$. It is a well known property of the 
QR-factorisation that it allows the solution of least squares
problems, i.e., given the overdetermined equation $\mathbf{G}h=f$, the 
``solution'' $h=\mathbf{R^{-1}Q^*}f$ has the property that it
minimizes $||\mathbf{G}h-f||^2$. Thus, e.g., if f is in the image of
$\mathbf{G}$ then $h$ is the (unique, if $\mathbf{R}$ is invertible)
vector in $\cT_M$ for which the equation holds. But if $f$ is not in
the image, then $h$ will be that vector in $\cT_M$ whose image is closest
to $f$, i.e., for which $f-\mathbf{G}h$ is orthogonal to the
image. Therefore, the QR-factorisation serves to compute the
``generalized inverse'' or Moore-Penrose inverse of a matrix
\cite{Penrose-1955}.

Now we can find the solution of the division problem as follows. Given
a smooth function $f$ on the interval $[-1,1]$ we compute its
expansion into Chebyshev polynomials up to degree $M$. Then we can
consider $f$ to be an element of $\cT_M$. We may also regard it as
being in $\cT_{2M}$ by taking the coefficients of the polynomials with
degree higher than $M$ to be zero. We also expand the divisor $g$,
another smooth function on the above interval and from its expansion
coefficients we construct the matrix $\mathbf{G}$. Then we compute the
QR-factorisation of $\mathbf{G}$ by standard methods (we take
Householder reflections). The solution $(f/g)$ is obtained by
computing $y=\mathbf{Q}^*f$, solving $\mathbf{R}y=z$ by back
substitution and, finally, by transforming back from the expansion
coefficients contained in the vector $z$ to the function $(f/g)$.

\begin{figure}[htbp]
  \begin{center}
    \makebox[\hsize]{\hfill\psfig{file=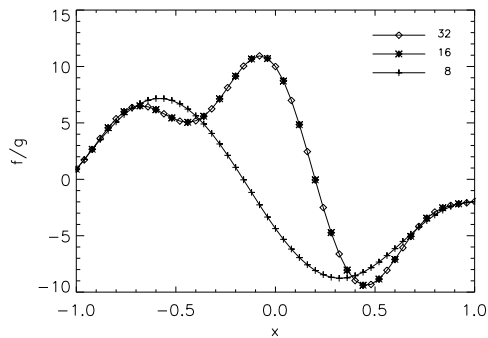 ,height=5cm}
      \hfill}
    \caption{The quotient (f/g) (see text)}
    \label{fig:fg}
  \end{center}
\end{figure}

\begin{figure}[htbp]
      \makebox[\hsize]{\hfill\psfig{file=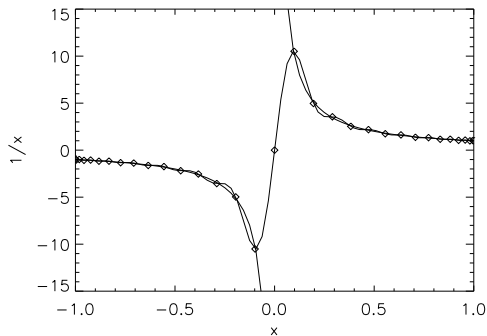,height=5cm}\hfill} 
  \caption{The quotient $1/x$ (see text)}
  \label{fig:inverse}
\end{figure}

Let us illustrate the above mentioned behaviour by an example. We take
$g(x)=x$ and $f(x) = \sin(10x)+2\cos(5x)-2$. Since $f(0)=0$, $f$ is
divisible by $g$ with $(f/g)(0)=5$. In Fig.~\ref{fig:fg} is shown the
exact quotient (solid line) and three approximations for $h=(f/g)$
with $M=8,16,32$ obtained by the procedure given above. Obviously, the
lowest approximation with $M=8$ is not usable. The reason for this is
that there is too much structure present in $f$ to be resolved with
only $8$ polynomials. Following the discussion in
\cite{KreissOliger-1972} and \cite{GottliebOrszag-1977}, we find that
in this case we need about $10$ polynomials to have enough resolution
power, and indeed, the approximation with $16$ polynomials is almost
indistinguishable graphically from the exact function. With $32$
polynomials the residual $||f-h\cdot g||_{\infty}$ is on the level of the
machine precision, see Table~\ref{tab:1}.
\begin{table}[htbp]
\[
    \begin{array}{|r|r|}
      M&||f-h\cdot g||_{\infty}\\
      \hline
      8&1.64\\
      16&2.19(-3)\\
      32&3.12(-14)\\
    \end{array}\]
    \parbox{10cm}{\caption{Maximum residual for approximations with
        different number $M$ of polynomials.}} 
    \label{tab:1}
\end{table}
To illustrate the behaviour when $f$ is not in the image of $g$, we
take $f(x)=1$ and $g(x)=x$. The result is shown in
Fig.~\ref{fig:inverse} for $M=32$.
The thin line is the exact function $1/x$ while the thick line is the
computed approximation $h$. The markers indicate the values at the
collocation points $x_k=-\cos(k\pi/M)$, where the approximation is
very good. However, inbetween the collocation points the agreement is
bad because of the high-frequency oscillation. The maximal residual is
in this case $||1-x\cdot h||_\infty=1.0$, which is the value of $f$ at
the zero of $g$. But the maximum value of $h$ is much higher,
$||h||_\infty=10.52$, one order of magnitude above the value of $f$ at
the zero of $g$.

In this case the zero was in the center of the interval. If, instead,
$g$ vanishes at the boundary, then the singular behaviour is much more
prominent. This is due to the clustering of the collocation points
towards the end of the interval which normally is a benefit, because
it allows a much more accurate approximation of functions there. In
our case it means a good approximation of the singular behaviour. If we
take $f(x)=1$ but $g(x)=1-x$ then we obtain a quotient with a maximum
value of $373.9$, more than two orders of magnitude above the value of
$f$ at the zero of $g$.

If we compute $(f/g)$ numerically from a function $f$ which is known
analytically to have common zeros with $g$, but has been obtained by
numerical means then $f$ will in general not vanish exactly at the
zeros of $g$. At those points its value $\eps$ will depend on the
accuracy of the algorithm used to compute $f$. Thus, we will have
$f(x)=g(x)h(x)+\eps$ (assuming for the moment that $g$ has only one
zero). Then, using the procedure described above, we would obtain $h$
together with some additional ``singular'' or ``high-frequency'' part,
which contaminates the result because its maximum value can be several
orders of magnitude beyond the value of $\eps$. The reason for this is
that $f$ is projected orthogonally onto the image of $g$. But we are
not so much interested in the orthogonal projection but rather we seek
a projection which annihilates that high-frequency part.

This can be achieved by a simple alteration of the projector
$\mathbf{P = QQ^*}$ in the following way. We seek a projector
$\mathbf{\tilde P}$ which has the same image as $\mathbf{P}$ but which 
annihilates a vector, say $y$. We make the ansatz
\begin{equation}
  \label{4.3}
  \mathbf{\tilde P} z = \mathbf{P}z-\scalprod{a}{z} \mathbf{P}y,
\end{equation}
where $a$ is some vector with $\scalprod{a}{y}=1$. For this to be a
projector we need in addition that $\scalprod{a}{\mathbf{P}z}=0$ for
all $z$. Thus, $a$ is orthogonal to the image and the simplest choice
for it is
\begin{equation}
  \label{4.4}
  a=\frac{y-\mathbf{P}y}{||y-\mathbf{P}y||^2}.
\end{equation}
Inserting the expression for $\mathbf{P=QQ^*}$, one readily finds
\[
  \mathbf{\tilde P} z = \mathbf{Q}\left[
    \mathbf{Q^*} z - \frac{\scalprod{y}{z} -
    \scalprod{\mathbf{Q^*}y}{\mathbf{Q^*} z}}{\scalprod{y}{y} -
    \scalprod{\mathbf{Q^*}y}{\mathbf{Q^*} y}}\,\mathbf{Q^*}y
  \right]
\]
Now it is clear what to do numerically. First the vector $y$ is
determined as the expansion coefficients of $1$ with respect to the
normalised Chebyshev polynomials and, once the QR-factors of
$\mathbf{G}$ have been found, the denominator of the factor above is
computed. Then each dividend is Chebyshev expanded to get the vector
$z$ of its expansion coefficients, the term in brackets above is
computed and, finally, the solution is obtained by inverting
$\mathbf{R}$. If this procedure is applied to find $1/x$ as before,
then one obtains exactly zero. And in case one tries to find
$(f(x)+\eps)/g(x)$ the result is the same no matter what value of
$\eps$.

So far, this procedure works only for the case where $g$ vanishes at a
single point. If there are more zeroes of $g$ present then the
procedure has to be modified. This modification is
straightforward. For two vectors $y_1$, $y_2$ to be annihilated, we
have the altered projection
\begin{align}
  \mathbf{\tilde P} z = \mathbf{P} z &+ \frac1{V_{12}} \left(
    \scalprod{y_2^\perp}{z} \scalprod{y_2^\perp}{y_1^\perp} 
    - \scalprod{y_1^\perp}{z} \scalprod{y_2^\perp}{y_2^\perp} \right) 
  \,\mathbf{P}y_1\nonumber \\
  &\label{4.5} + \frac1{V_{12}} \left(
    \scalprod{y_1^\perp}{z} \scalprod{y_1^\perp}{y_2^\perp} 
    - \scalprod{y_2^\perp}{z} \scalprod{y_1^\perp}{y_1^\perp} \right) 
  \,\mathbf{P}y_2,
\end{align}
where $V_{12} = \scalprod{y_1^\perp}{y_1^\perp}
\scalprod{y_2^\perp}{y_2^\perp} - \scalprod{y_2^\perp}{y_1^\perp}^2$
and where $y^\perp = y-\mathbf{P}y$ is the part of $y$, orthogonal to
the image of $\mathbf{P}$. In principle, this can be generalized to
even more $y$'s but the formulae become more and more complicated. The
meaning of the vectors $y_i$ is the following. Suppose $g$ has $n$
zeros in the interval. Then it is described by a polynomial of degree
$n$. Each function in the image of $\mathbf{G}$ is then necessarily a
polynomial of degree at least $n$. Therefore, the projection onto the
image of $\mathbf{G}$ has to annihilate all the lower degree
polynomials. This implies that one can take the standard basis vectors
$(1,0,\ldots)$, $(0,1,0\ldots)$,\ldots for the vectors $y$.

Let us now describe how to implement this method of division. We
assume that a solution $\phi$ of the Yamabe equation has been
obtained. Together with the boundary defining function $\omega$ it
defines the conformal factor $\Omega = \omega \phi^{-2}$. Since $\phi
> 0$, both $\Omega$ and $\omega$ vanish at exactly the same
points. From $\phi$ and the geometry of $\Sigma$ one determines the
components of the tensor fields in \eqref{2.5} and \eqref{2.6} by
differentiation and algebraic manipulations. Let $\psi$ be any one of
those components. Analytically, one knows that $\psi$ shares the same
zeroes as $\Omega$ if the free data have been specified appropriately,
so that, ideally, there is no problem when dividing $\psi$ by
$\Omega$. However, the values of $\psi$ will never be exactly zero at
the zeroes of $\Omega$ and we have the situation described above.

We now assume for simplicity that $\omega$ is a function only of $v$,
which is the case in the first example in section ~\ref{sec:yam}. This
assumption is easily removed. To divide $\psi$ by $\Omega$ we divide
by $\omega$ and then multiply with $\phi^2$. The function $\psi$ is
represented by a two-dimensional array. To divide this array by
$\omega$ we divide it row by row, each row consisting of the values
$\psi_i=\psi(u_i)$ at the points of constant $u=u_i$. The boundary
function $\omega$ is Chebyshev-transformed and from its spectral
representation we construct the matrix $\mathbf{G}$ and its reduced
QR-factorisation. In the example with $\omega(v) = \frac12(1-v^2)$
there are only two expansion coefficients because $\omega$ is an even
quadratic polynomial. Then we compute the projection \eqref{4.5} of
the Chebyshev-transformed row $\psi_i$ onto the image of
$\mathbf{G}$. The two vectors $y_1$ and $y_2$ have components
$\delta_{k0}$ and $\delta_{k1}$ respectively, thus being the spectral
representation of the two lowest degree polynomials. The projection
corresponds to the subtraction from $\psi_i$ of an affine function
$av+b$ which agrees with $\psi_i$ at the zeroes $v=\pm1$ of
$\omega$. Thus, the projected function vanishes on the boundary and we
can divide it by inverting the QR-factors of $\mathbf{G}$. Performing
these operations on all rows of $\psi$ finally yields the result
$\psi/\omega$. 

\begin{figure}[htbp]
  \makebox[\hsize]{\hfill\psfig{file=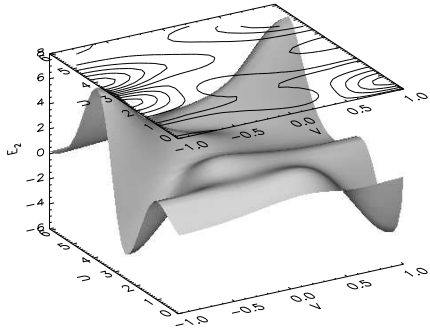,height=5cm}\hfill}
  \caption{A component of the rescaled Weyl tensor}
  \label{fig:e2}
\end{figure}

As an example we show in Fig.~\ref{fig:e2} a component of the
rescaled Weyl tensor which involves two divisions by $\omega$, see
eqns. \eqref{2.5} and \eqref{2.6}. The
maximal residual is in this case $||\psi - E\cdot\omega||_{\infty}
\approx 4(-16)$ which is on the level of machine accuracy.
This function is obtained from the first solution of the Yamabe
equation \eqref{2.4} presented in the previous section. There are no
interior $\scri$'s and so the boundary of space-time coincides with
the boundary of the grid.

\section{Conclusion}
\label{sec:concl}

We have presented in this paper a way for obtaining hyperboloidal
initial data sets for the conformal Einstein equations by using
pseudo-spectral methods. It has been demonstrated that these can be
efficient and powerful tools also in numerical relativity. We have
presented some results obtained under certain simplifying
assumptions. In particular, these are the dimensional reduction by
introducing a hypersurface orthogonal symmetry and the use of periodic
boundary conditions.

Although the results presented here are encouraging there are some
points to be stressed. One desirable thing for the evolution of these
initial data is that they be extended beyond $\scri$ in a way which is
as smooth as possible. This means that one has one or more $\scri$'s
in the interior of the grid just as in the second example presented in
section ~\ref{sec:yam} and then the boundary of the grid is outside
the physical space-time. If the Yamabe equation is not singular at the
grid boundary this means that we have to give some boundary
values. These in turn influence the solution up to the inner $\scri$'s
while inside these $\scri$'s the solution is determined from the
equation alone. In general, these two parts of the solution will not
match smoothly across the inner boundaries, there will be some higher
derivative which jumps. This is due to the fact that the third
one-sided normal derivative of the solution on $\scri$ is
characterised by the mass aspect of that part of the region on which
the derivative is taken \cite{HF-priv}. And since there is no obvious
relation between the values of the solution on the grid boundaries and
the mass aspect on $\scri$ there is no guarantee that the third
derivatives of the solution taken on either side of $\scri$ agree on
$\scri$. This implies that the initial data, and most notably the Weyl
curvature, are not as smooth across $\scri$ as they should be.

There might be a possibility, though, to overcome this problem (due to
S. Brandt \cite{Brandt}). If we keep the grid boundaries singular then
we do not need to specify boundary conditions there. Then it might be
possible that the specification of free data which are analytic
enforces the equality of the mass aspects of the different regions on
their common $\scri$. This possibility needs to be studied in detail.

We have shown here some results in two dimensions. The generalisation
of the Yamabe solver to full three dimensions should be
straightforward. The situation is not so clear for the divison
part. The fact that the two-dimensional division is performed by
stacking together one-dimensional divisions might be impractical in
three dimensions. 

Another limitation of the division procedure is the fact that it has
to be changed whenever the boundary defining function acquires more
zeros. This is more a matter of practicality than of principle. A
final remark concerns the accuracy of the division method.  Although
each individual division can be made quite accurately with this method
this does not imply that the quotients are similarly accurate, in the
sense that the conformal constraint equations are satisfied to any
accuracy comparable to that of the divisions. The reason for this is
that the division process only ``cures the symptoms'', namely the mild
non-vanishing of the dividends on the boundary. It does not remove the
reason for this phenomenon. That would probably be related to the fact
that the Yamabe equation does not only enforce the behaviour of the
function on the boundary but also of its first derivative. This has
not been used so far in the solution process of the
equation. Therefore, we cannot expect that the derivative of the
numerical solution has the values it should have on the boundary. And
so the constraints are only satisfied to the degree with which these
values are attained. This problem is currently being studied.

Thus, to summarize, we feel that the Yamabe solver can be made a
rather efficient and accurate tool, while the divisor may have its
limitations. These come mostly from the fact that it cannot be easily
applied for more general boundary functions. There is a completely
different approach to the division problem developed by
H\"ubner~\cite{Huebner-1998-3}, where one solves a linear elliptic
equation for the quotient $\psi/\omega$ which is singular at
$\omega=0$. 

The results and methods presented in this paper demonstrate the
feasibility of numerically determining initial data sets for the
conformal field equations along the lines of
\cite{AnderssonChrusciel-1994,AnderssonChruscielFriedrich-1992}. They
enable the numerical evolution of general relativistic space-times
using the ``conformal method''.

\section*{Acknowledgments}

It is a pleasure to thank the members of the mathematical relativity
section of the MPI f\"ur Gravitationsphysik in Potsdam. I am particularly
grateful to H. Friedrich and P. H\"ubner for useful discussions on the
analytical and numerical intricacies of the Yamabe problem. Furthermore,
I wish to thank S. Bonazzola for letting me have a look into his
pseudo-spectral hydro-code where I found some very useful tricks and
techniques. This work is funded by the DFG grant FR-848/3.


\end{document}